\documentclass[twocolumn,showpacs,preprintnumbers,amsmath,amssymb,prb]{revtex4}
\usepackage{graphicx}
\usepackage{dcolumn}
\usepackage{bm}
\usepackage{subfigure}
\usepackage{float}
\usepackage{color}

\begin{document}

\title{Phase transitions in 2D $J_{1}-J_{2}$ model with arbitrary signs of
exchange interactions}
\author{A.V. Miheyenkov$^{+*}$, A.V. Shvartsberg$^{*}$, A.F. Barabanov$^{+}$}
\affiliation{$^+$Institute for High Pressure Physics RAS, 142190 Moscow (Troitsk), Russia%
\\
$^*$Moscow Institute of Physics and Technology, 141700 Dolgoprudny, Russia}
\date{\today }

\begin{abstract}
The ground state of the $S=1/2$ $J_{1}-J_{1}$ Heisenberg model on the 2D
square lattice with arbitrary signs of exchange constants is considered.
States with different spin long-range order types (antiferromagnetic
checkerboard, stripe, collinear ferromagnetic) as well as disordered
spin-liquid states are described in the frames of one and the same
analytical approach. It is shown inter alia, that the phase transition
between ferromagnetic spin liquid and long-range order ferromagnet is a
second-order one. On the ordered side of \ the transition the ferromagnetic
state with rapidly varying condensate function is detected.
\end{abstract}

\maketitle

Investigation of the two-dimensional frustrated Heisenberg model is of
current importance for understanding magnetic properties of various layered
compounds. Spin subsystem of $CuO_{2}$ planes in cuprate high-temperature
superconductors (HTSC) can be described by $J_{1}-J_{2}$ Heisenberg model on
the square lattice with spin $S=1/2$ and antiferromagnetic signs of both
exchange constants. Intensively studied layered vanadium oxides can be
described in the frames of same model, but not only with antiferromagnetic
exchanges.

\begin{figure}[bp]
\begin{center}
\includegraphics[width=8.0cm]{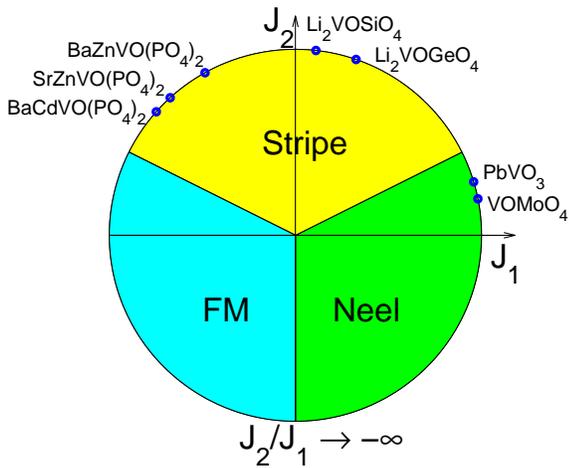}
\caption{Phase diagram of the $J_1$--$J_2$ Heisenberg model on the 2D square
lattice in the classical limit. Dots represent the relations between $J_1$
and $J_2$ for the better-studied vanadates (data from \cite{Nath08,Tsirlin09}). }
\label{fig1}
\end{center}
\end{figure}

In the classical limit $S\gg 1$ at zero temperature three types of
long-range order (LRO) are realized: ferromagnetic (FM), Neel
antiferromagnetic (AFM) and columnar (stripe). At the points $%
J_{2}/\left\vert J_{1}\right\vert =0.5$ there are first order phase
transitions from checkerboard AFM\ order to stripe for $J_{1}>0$ and from
stripe to ferromagnetic order for $J_{1}<0$, at point $J_{1}=0$, $J_{2}=-1$
there is a transition from AFM to FM order. The positions of the
better-studied vanadates on the classical $J_{1}$--$J_{2}$ model phase
diagram are shown in Fig.~\ref{fig1} (the data from Refs.~\onlinecite{Nath08,Tsirlin09}%
).

At $T\neq0$ long-range order due to Mermin-Wagner theorem is impossible
for any spin, at $T=0$ for large $S$ LRO exists throughout
the "$J_{1}-J_{2}$-circle". Nevertheless, it is generally accepted that for $S=1/2$
even for $T=0$ spin fluctuations near phase transition points
lead the system to one of the singlet states without LRO and
with nonzero spin gap. The structure of disordered phases remains debatable.
Usually the following states are considered: spin liquid, conserving
translational and $SU(2)$ symmetry of the Hamiltonian; plaquette lattice covering,
which breaks translational symmetry, but conserves $SU(2)$
symmetry; and states that break both translational and $SU(2)$ symmetry.

In the present work the ground state of 2D $J_{1}-J_{2}$ Heisenberg model is
investigated in the frames of spherically symmetric self-consistent approach
(SSSA) for two-time retarded Green's functions (Refs.~%
\onlinecite{Shimahara91,BarBer94-both}, see also recent review in Ref.~%
\onlinecite{BarMixTMF11}). This approach automatically conserves $SU(2)$
symmetry of the Hamiltonian, translational symmetry and spin constraint on
the site. Unlike previous treatments of $S=1/2$ model, we
investigate the entire phase diagram for arbitrary values of $J_{1}$ and
$J_{2}$, including cases of $J_{1}<0$, $J_{2}>0$ and $J_{1}<0$, $J_{2}<0$.

In the quantum limit $S=1/2$, the first quadrant of the diagram
$0\leq\varphi\leq\pi/2$, $\tan\varphi=J_{2}/J_{1}$, $J_{1},\ J_{2}>0$, has
been studied up to now most extensively. In this case a disordered state
(spin liquid) appears between AFM and stripe LRO phases. The transitions
AFM --- spin liquid --- stripe phase are continuous in the frames of SSSA.

The region $J_{1}<0$, $J_{2}>0$ for $S=1/2$ of the phase diagram has been
investigated in frames of SSSA in \cite{Hartel11,Hartel13}, where the
the first order transition between FM LRO state and spin liquid
has been stated. As it will be seen hereafter, unlike \cite{Hartel11,Hartel13},
our consideration leads a continuous second-order transition between the mentioned
states, the properties of FM state being significantly modified near the transition.

Before discussing the phase diagram in the whole range of the angle $\varphi$,
let us write down the Hamiltonian $H$ and the form of spin-spin Green's
function $G^{zz}\left(\omega,\mathbf{q}\right)$, which can be obtained in
the frames of SSSA \cite{Shimahara91,BarBer94-both,BarMixTMF11,Hartel11,Hartel13}
(for SSSA $G^{zz}=G^{xx}=G^{yy}$; $\langle S_{\mathbf{i}}^{\alpha}\rangle=0,$
$\alpha=x,y,z$).
\begin{equation}
H=\frac{J_{1}}{2}\sum_{\mathbf{i},\mathbf{g}}\widehat{\mathbf{S}}_{\mathbf{i}%
}\widehat{\mathbf{S}}_ {\mathbf{i}+\mathbf{g}}+\frac{J_{2}}{2}\sum_{\mathbf{i%
},\mathbf{d}}\widehat{\mathbf{S}}_{\mathbf{i}} \widehat{\mathbf{S}}_{\mathbf{%
i}+\mathbf{d}}  \label{Ham}
\end{equation} \begin{equation}
G^{zz}\left(\omega,\mathbf{q}\right)=\langle S_{\mathbf{q}}^{z}|S_{-\mathbf{q%
}}^{z}\rangle_{\omega}=\frac{F_{\mathbf{q}}}{\omega^{2}-\omega_ {\mathbf{q}%
}^{2}}  \label{Gzz}
\end{equation} \begin{equation}
F_{\mathbf{q}}=-8[J_{1}(1-\gamma_{g})c_{g}+J_{2}(1-\gamma_{d})c_{d}]
\label{Fq}
\end{equation} \begin{equation}
\gamma_{g}(\mathbf{q})=\frac{1}{z_{g}}\sum_{\mathbf{g}}e^{i\mathbf{qg}}=%
\frac{1}{2} (\cos(q_{x})+\cos(q_{y}))
\end{equation} \begin{equation}
\gamma_{d}(\mathbf{q})=\frac{1}{z_{d}}\sum_{\mathbf{d}}e^{i\mathbf{qd}%
}=\cos(q_{x})\cos(q_{y})
\end{equation}
where $\mathbf{g},\ \mathbf{d}$ are vectors of nearest and next-nearest
neighbors, $c_{R}=\langle S_{\mathbf{n}}^{z}S_{\mathbf{n+R}}^{z}\rangle$ ---
spin-spin correlation functions on the corresponding coordination spheres, $%
z_{g}=z_{d}=4$ --- number of sites on the first and the second coordination
spheres. Hereafter all the energetical parameters are set in the units of
$J=\sqrt{J_{1}^{2}+J_{2}^{2}}$.

For further analysis, it is convenient to represent the spin excitation
spectrum $\omega_{\mathbf{q}}^{2}$ (\cite{BarBer94-both,BarMixTMF11}) in three
following forms:
\begin{eqnarray}
\omega_{\mathbf{q}}^{2} & = & 2A\left(\mathbf{q}\right)\left(1-\gamma_{g}%
\right)\left(1-\gamma_{g}+\delta_{FM} \left(\mathbf{q}\right)\right)=
\nonumber \\
& = & -2A\left(\mathbf{q}\right)\left(1-\gamma_{g}\right)\left(1+\gamma_{g}+%
\delta_{AFM} \left(\mathbf{q} \right)\right)=  \nonumber \\
& = & -2A\left(\mathbf{q}\right)\left(1-\gamma_{g}\right)\left(1+\gamma_{d}+%
\delta_{Stripe} \left(\mathbf{q}\right)\right)  \label{eq:wq2}
\end{eqnarray}

Expressions for $A$ and $\delta$ from (\ref{eq:wq2}) are rather unwieldy,
and we do not present them completely. We will just present the
form of $A\delta_{AFM}$ as an example:
\begin{equation*}
A\,\delta _{FM} =8J_{1}J_{2}\left( \widetilde{c}_{dg}\!-\!\widetilde{c}%
_{g}\right)+\!J_{1}^{2}\left(1\!-\!20\widetilde{c}_{g}\!+\!8\widetilde{c}_{d}\!+\!4\widetilde{c}%
_{2g}\right)+
\end{equation*} \begin{equation}
+\frac{1\!-\!\gamma _{d}}{1\!-\!\gamma _{g}}[8J_{2}J_{1}\left( \widetilde{c}%
_{dg}\!-\!\widetilde{c}_{g}\right) +\!J_{2}^{2}\left(1\!-\!20\widetilde{c}%
_{d}\!+\!8\widetilde{c}_{2g}\!+\!4\widetilde{c}_{2d}\right) ]  \label{del_FM}
\end{equation}

In (\ref{del_FM}) the correlators $\widetilde{c}_{r}=\alpha c_{r}$ are
written in one vertex $\alpha$ approximation (\cite{BarMixTMF11,Hartel11}).
Five correlators $c_{r}$ ($r=g,\ d,\ 2g,\ r_{gd}=|\mathbf{g}+\mathbf{d}|,\ 2d$)
and vertex correction $\alpha$ are obtained self-consistently through the
Green's function $G^{zz}$. The additional condition is the exact sum rule fulfillment $\langle\widehat{\mathbf{S}}_{\mathbf{i}}^{2}\rangle=3/4$.

The introduced parameters $\delta_{FM}\left(\mathbf{q}\right)$,
$\delta_{AFM}\left(\mathbf{q}\right)$, and $\delta_{Stripe}\left(\mathbf{q}\right)$
have a clear physical meaning and define the spin excitation spectrum basic properties.
For all phases --- three ordered (AFM, stripe, and FM), and spin
liquid --- the spin gap is closed at the zero point $\mathbf{\Gamma}=\left(0,0\right)$.
In the FM phase the spectrum around $\mathbf{\Gamma}$ is
quadratic in $q$, for other phases it is linear. Near the transitions
to FM from the neighboring phases the spectrum around $\mathbf{\Gamma}$ has
the form $\omega_{q}\sim q\sqrt{\delta_{FM}+\frac{q^{2}}{4}}$. So $\delta_{FM}$
dictates the conversion from FM spectrum $\omega_{q}\sim q^{2}$
to $\omega_{q}\sim q$. In the AFM phase the spin gap is closed not only in
$\mathbf{\Gamma}$, but also in AFM point $\mathbf{Q}=\left(\pi,\pi\right)$.
When approaching to the AFM phase from the neighboring phases the spectrum
around $\mathbf{Q}$ is $\omega_{q}\sim\sqrt{\delta_{AFM}+\varkappa^{2}}$,
$\varkappa=|\mathbf{Q-q}|$, i.e. $\delta_{AFM}$ directly defines the gap
$\Delta_{AFM}$ in the spectrum. For the stripe phase and it's neighborhood
the situation is similar to that for AFM phase (with corresponding substitutions,
the role of control point goes from $\mathbf{Q}$ to to stripe points $\mathbf{X}=(0,\pi),(\pi,0)$).

Thus, vanishing of any of the three parameters $\delta_{FM}$, $\delta_{AFM}$,
$\delta_{Stripe}$ defines transition to the corresponding ordered phase
and simultaneous alteration of the spectrum near the corresponding
control point (transition from linear to quadratic for FM and vanishing of
the spin gap in the Dirac spectrum for two others). For the spin liquid the
spectrum gap is opened in the whole Brillouin zone except $\mathbf{\Gamma}$.

Let us depict in more detail the description of the spin LRO. The
structure factor has the form
\begin{eqnarray}
c_{\mathbf{q}} & = & \langle S_{\mathbf{q}}^{z}S_{-\mathbf{q}}^{z}\rangle=-%
\frac{1}{\pi}\int d\omega \,n\left(\omega_{\mathbf{q}}\right)\mathrm{Im}G^{zz}
\left(\omega,\mathbf{q}\right)=
\nonumber \\
& = & \frac{F_{\mathbf{q}}}{2\omega_{\mathbf{q}}}\left(2n\left(\omega_{%
\mathbf{q}}\right)+1\right)
\end{eqnarray}
here $n\left(\omega_{\mathbf{q}}\right)$ is Bose function. Correlation
functions are expressed through the structure factor as
\begin{eqnarray}
c_{R} & = & \langle S_{\mathbf{n}}^{z}S_{\mathbf{n+R}}^{z}\rangle=\sum_{%
\mathbf{q}}c_{\mathbf{q}}e^{i\mathbf{qR}}=  \nonumber \\
& = & c_{cond}\sum_{\mathbf{q}_{0}}e^{i\mathbf{q}_{0}\mathbf{R}}+\frac{1}{%
4\pi^{2}}\int d\mathbf{q}e^{i\mathbf{qR}}\frac{F_{\mathbf{q}}}{2\omega_{%
\mathbf{q}}}  \label{eq:Cr}
\end{eqnarray}
where the condensate part is
\begin{equation}
c_{cond}=\lim_{T\rightarrow0}\frac{1}{4\pi^{2}}\int d\mathbf{q}%
n\left(\omega_{\mathbf{q}}\right)\frac{F_{\mathbf{q}}}{\omega_{\mathbf{q}}}
\end{equation}

At $T\rightarrow0$ $\delta$-like peaks in the structure factor can appear at
some points $\mathbf{q}_{0}$ of the Brillouin zone (where $\omega_{\mathbf{q}}$
tends to zero), this peaks being induced by the Bose function
$n\left(\omega_{\mathbf{q}}\right)$. Then the condensate term
$c_{cond}$ appears in the correlation functions $c_{R}$. This
corresponds to the LRO existence ($c_{cond}$ defines spin-spin
correlator at the infinity). The term without $n\left(\omega_{\mathbf{q}}\right)$
on the right hand side of (\ref{eq:Cr})
goes to zero as $R\rightarrow\infty$.

For AFM and stripe long-range orders the condensate term appears as the
spectrum $\omega_{\mathbf{q}}$ vanishes correspondingly at the points
$\mathbf{Q}$ and $\mathbf{X}$. As mentioned above, the spectrum near this
points is (in the corresponding phase) $\omega_{\mathbf{q}}\sim\varkappa$,
where $\varkappa=\left\vert \mathbf{q}-\mathbf{Q}\right\vert$ or
$\left\vert \mathbf{q}-\mathbf{X}\right\vert $. The
Green's function numerator $F_{\mathbf{q}}$ does not
vanish at this points. The spectrum linearity and nonzero $F_{\mathbf{q}}$ value
constitute the condition for condensate to appear \cite{Shimahara91}.

In the presence of FM LRO spin condensate appears at the point
$\mathbf{\Gamma}$. Near this point the Green's function numerator
$F_{\mathbf{q}}\sim q^{2}$, so the spectrum near $\mathbf{\Gamma}$
is to be also quadratic $\omega_{\mathbf{q}}\sim q^{2}$ for the condensate
to appear.

Note that if the third exchange interaction $J_{3}$ is added to the
model, the helical LRO can also be realized. In the $J_{1}-J_{2}-J_{3}$ model
the condensate peak point in the structure factor can be located not only
at $\mathbf{\Gamma}$, $\mathbf{Q}$, or $\mathbf{X}$, but also at arbitrary
incommensurate point on the side or diagonal of the Brillouin zone \cite{MixBarJETPL11}.

\begin{figure}[tp]
\begin{center}
\includegraphics[width=8.8cm]{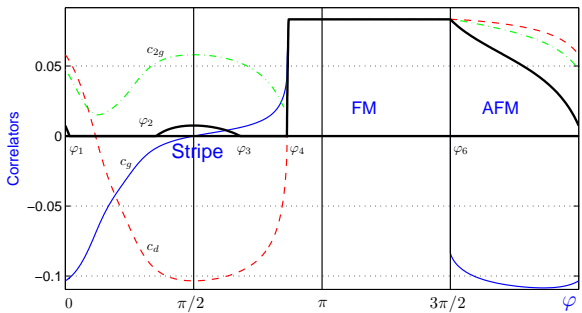}
\end{center}
\caption{The condensate $c_{cond}$ (absolute value of spin-spin
correlator at infinity) and spin-spin correlators on the first three
coordination spheres dependence on $\varphi$ ($J_1=\cos
\varphi,$ $J_2=\sin\varphi$). Black bold line shows $c_{cond}$, blue solid
line --- $c_g$, red dotted --- $c_d$, green dash dotted --- $c_{2g}$. The points of
phase transitions are marked on the x-axis: $\varphi_1$ -- AFM $\rightarrow$
SL$^1$ transition, $\varphi_2$ --- SL$^1$ $\rightarrow$ Stripe,
$\varphi_3$ --- Stripe $\rightarrow$ SL$^2$, $\varphi_4$ ---
SL$^2$ $\rightarrow$ FM$^1$ (for the rescaled vicinity of $\varphi_4$ see
Fig.4), $\varphi_6$ --- FM$^2$ $\rightarrow$ AFM transition. See
text for details.}
\label{fig2}
\end{figure}

\begin{figure}[bp]
\begin{center}
\includegraphics[width=8.8cm]{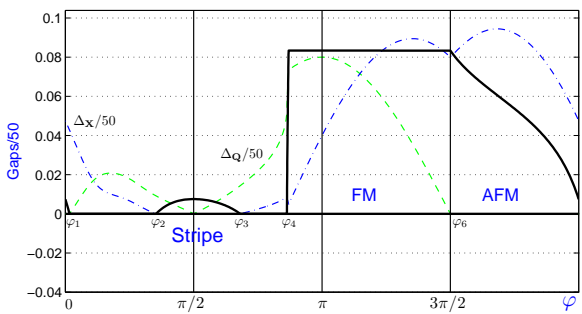}
\end{center}
\caption{Spin gaps $\Delta_\mathbf{Q}$ and $\Delta_\mathbf{X}$
($\Delta_\mathbf{Q}$/50 and $\Delta_\mathbf{X}$/50 are shown)
at the points $\mathbf{Q}=(\pi,\pi)$ (green dashed line) and
$\mathbf{X}=(0,\pi),(\pi,0)$ (blue dash-dotted) of the Brillouin zone
as functions of $\varphi$ ($J_1=\cos\varphi,$ $J_2=\sin\varphi$).
Black solid line --- condensate $c_{cond}$. All the points $\varphi_1$--$\varphi_6$ are
the same as in Fig.2.}
\label{fig3}
\end{figure}

Fig.~\ref{fig2} shows the phase diagram at $T\rightarrow0$, the
condensates and correlators corresponding to first three coordination
spheres being depicted. Fig.~\ref{fig3} represents spin gaps in the symmetrical points.
In the interval $0\leq\varphi\leq\varphi_{1}=0.051$ AFM LRO is realized: spin gap at AFM
point $\mathbf{Q}$ is zero, $\Delta_{\mathbf{Q}}=0$, the spectrum near
$\mathbf{Q}$ is linear in $|\mathbf{q}-\mathbf{Q}|$, there is a nonzero AFM
condensate $c_{cond}^{AFM}$.

At $\varphi=\varphi_{1}$ condensate $c_{cond}^{AFM}$ vanishes, AFM gap
$\Delta_{\mathbf{Q}}$ opens, and the spectrum becomes ungapped in the whole
Brillouine zone, except trivial zero point $\mathbf{\Gamma}$, where it
remains linear. The system turns to spin liquid state (let's denote it by
SL$^1$), which is realized in the interval $\varphi_{1}\leq\varphi\leq\varphi_{2}=1.111$.
In this phase LRO is absent, and short-range order transforms
with growing $\varphi$ from the AFM-like ($c_{g}<0$, $|c_{g}|>c_{d}>c_{2g}>0$)
to the one typical for stripe phase ($c_{d}<0,$ $c_{2g}>0$, $|c_{d}|>c_{2g}>|c_{g}|$).
At the same time spin gap at the point $\mathbf{Q}$ passes through the maximum, and the gap at stripe points $\mathbf{X}$ monotonically decreases (Fig.~\ref{fig3}).

At $\varphi=\varphi_{2}$ the stripe gap $\Delta_{\mathbf{X}}$ vanishes, the
spectrum at stripe points $\mathbf{X}$ becomes linear, and condensate
$c_{cond}^{Stripe}$ becomes nonzero, the system turns to the LRO stripe phase,
which is realized in the interval $\varphi_{2}\leq\varphi\leq\varphi_{3}=2.141$.
Note the very interesting point $\varphi=\pi/2$, where $J_{1}=0$, $J_{2}=1$.
The lattice with this exchange couplings splits into two
noninteracting AFM sublattices. Then it is obvious that
$c_{d}\left(\pi/2\right)=c_{g}\left(0\right)$,
$c_{2g}\left(\pi/2\right)=c_{d}\left(0\right)$ (see Fig.~\ref{fig2}).
Note that at $\varphi=\pi/2$, as in AFM phase, $\Delta_{\mathbf{Q}}=0$ (see Fig.~\ref{fig3}),
however, this does not lead to AFM LRO, because
$F\left(\mathbf{Q},\varphi=\pi/2\right)$ also vanishes, and as a result
$c_{cond}^{AFM}\left(\varphi=\pi/2\right)=0$.

At the point $\varphi=\varphi_{3}$ stripe gap $\Delta_{\mathbf{X}}$ opens,
and the system again turns to the spin liquid state (SL$^{2}$),
realized in the interval $\varphi_{3}<\varphi<\varphi_{4}=2.712$ (but
the structure of the short-range order differs from that at
$\varphi_{1}\leq\varphi\leq\varphi_{2}$). It is worth noting, that
the next-nearest neighbor correlator $c_{d}$ remains negative throughout
the SL$^{2}$ existence, i.e. the short-range order is not rearranged to the
FM-like, where $c_{g},c_{d},c_{2g}>0$. The absolute value of $c_{d}$ almost
everywhere, except tiny region near $\varphi_{4}$, is larger than the
nearest neighbor correlator $c_{g}$.

Let us emphasize ones more, that for all the mentioned phases the spectrum near
$\mathbf{\Gamma}=(0,0)$ is linear in $q$.

\begin{figure}[tb]
\begin{center}
\includegraphics[width=8.8cm]{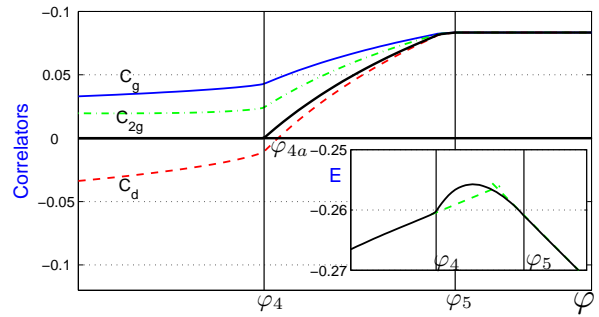}
\end{center}
\caption{Condensate $c_{cond}$ and correlators $c_{g}$,
$c_{d}$, $c_{2g}$ evolution from spin liquid SL$^{2}$ to ferromagnetic state FM$^{2}$.
As in Fig.2, black bold line --- $c_{cond}$, blue solid --- $c_{g}$, red dotted --- $c_{d}$,
green dash dotted --- $c_{2g}$.
$\varphi _{4}$ corresponds to the transition from spin liquid $SL^{2}$ to
ferromagnetic state FM$^{1}$, in the narrow region $\varphi _{4}$--$\varphi _{4a}$
short-range FM order is absent, while long-range FM order is present (see text),
$\varphi _{5}$ --- the border between ferromagnetic regions FM$^{1}$ and FM$^{2}$ (see text).
\newline
Inset: black solid line --- energy per site $\varepsilon $ of  the present work,
green dashed lines --- energy extrapolation for the solutions SL$^{2}$ and FM$^{2}$ from
Ref.~\onlinecite{Hartel11}. The intersection corresponds to first-order
transition between spin liquid and ferromagnet, stated in \cite{Hartel11}.
}
\label{fig4}
\end{figure}

At $\varphi=\varphi_{4}$ there again appears a phase with LRO
(ferromagnetic) and nonzero corresponding condensate $c_{cond}^{FM}$.
Spectrum near the point $\mathbf{\Gamma}$ becomes quadratic in $q$ (and the gap
$\Delta_{\mathbf{\Gamma}}(\varphi_{4})=0$). Fig.~\ref{fig2} and Fig.~\ref{fig3}
demonstrate, that two regions are distinguishable in this phase --- FM$^{1}$ and FM$^{2}$.
FM$^{1}$ covers in the tiny interval $\varphi_{4}<\varphi<\varphi_{5}=2.733$.
Here condensate $c_{cond}^{FM}$ grows rapidly with the increase of $\varphi$
from $c_{cond}^{FM}=0$ to the maximal value $c_{cond}^{FM}=1/12$.
Note, that near $\varphi_{4}$ FM LRO without
FM short-range order is realized, $c_{d}<0$ (the corresponding
interval is $\varphi_{4}<\varphi<\varphi_{4a}=2.713$). For
$\varphi\geq\varphi_{5}$ (FM$^{2}$ region) all the correlators and the condensate
are equal to $1/12$ and the vertex correction $\alpha=3/2$ \cite{Shimahara91,Hartel11}).

FM$^{1}$ region was not detected (Fig.~\ref{fig4}) in \cite{Hartel11}. The inset
of Fig.~\ref{fig4} shows the energy at the transitions SL$^{2}$
$\rightarrow$ FM$^{1}$ $\rightarrow$ FM$^{2}$. Dashed line
is the extrapolatin of SL$^{2}$ energy to the intersection with
the FM$^{2}$ energy (from \cite{Hartel11}). Based on
this extrapolation, it was concluded in \cite{Hartel11}, that a first order
transition occurs near the intersection point. Our consideration leads to
a conclusion (see Fig.~\ref{fig4}), that the energy derivative is continuous
between SL$^{2}$ and FM$^{1}$.

Note that standard FM$^{2}$ solution \cite{Shimahara91,Hartel11} exists also
for angles $\varphi<\varphi_{5}$, down to $\varphi=\pi-\arctan\left(1/2\right)$,
but in this region it happens to be metastable relative to FM$^{1}$
and SL$^{2}$.

At the angles $\varphi\geq\varphi_{5}$ the FM$^{2}$ solution is realized up
to $\varphi_{6}=3\pi/2$. This point is a very special one.
At $\varphi_{6}$ the lattice is splitted into two noninteracting
sublattices. At $\varphi\rightarrow\varphi_{6}-0$ there is no frustration with
respect to the FM order, at $\varphi\rightarrow\varphi_{6}+0$
--- no frustration with respect to the AFM order. Therefore it is
physically obvious that in the quantum limit a transition between
these two phases is of the first order, as do our calculations confirm.
Let us also note that, as it can be seen from Fig.~\ref{fig3}, at $\varphi\rightarrow\varphi_{6}+0$
($J_{1}=+0$, $J_{2}=-1$), AFM condensate, i.e. absolute value of the
spin-spin correlation function at infinity, is much larger than in the
"standard" AFM ($\varphi=0$, $J_{1}=1$, $J_{2}=0)$, and is equal to FM
condensate at $\varphi\rightarrow\varphi_{6}-0$. It means that FM
next-nearest neighbor exchange with zero nearest exchange leads to
stronger AFM order, than nearest AFM exchange with zero
next-nearest one.

\begin{figure}[tbp]
\begin{center}
\includegraphics[width=8.8cm]{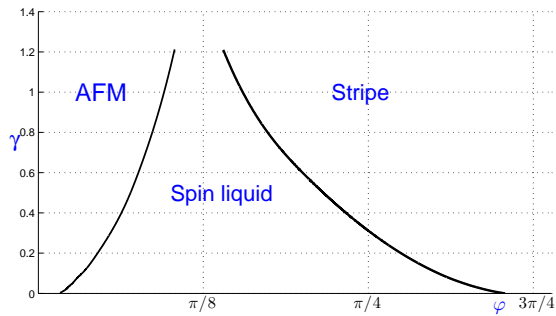}
\end{center}
\caption{Dependence of spin liquid SL$^1$ borders position on the damping
parameter $\protect\gamma$ (see text for details). }
\label{fig5}
\end{figure}

In conclusion let us note, that the significant spin excitations damping
can be expected near the transitions spin liquid $\rightarrow$ LRO
phase. Accounting for the damping can shift the boundary of the corresponding
transition. This is demonstrated in Fig.~\ref{fig5}, where the dependence of
SL$^{1}$ phase boundaries on the damping parameter $\gamma$ is represented.
We used the simple semiphenomenological approximation for the
Green's function $G_{\gamma}^{zz}$, conserving correct analytical properties
(see \cite{BarMixTMF11} for details).
\begin{equation}
G_{\gamma}^{zz}\left(\omega,\mathbf{q}\right)=\frac{F_{\mathbf{q}}}{%
\omega^{2}- \omega_{\mathbf{q}}^{2}+i\omega\gamma}
\end{equation}
It can be seen in Fig.~\ref{fig5}, that the SL phase boundaries are sensitive to the
value of damping. Nethertheless, our estimates show, that there are no
topological modifications of the phase diagram for any reasonable
values of damping.

To summarize, in the present work the entire phase diagram of the 2D
$J_{1}-J_{2}$ $S=1/2$ Heisenberg model is considered in the frames of
one and the same approach. It is shown, that the transitions between all
ordered and disordered phases are continuous, except the transition
FM$\rightarrow$AFM at $J_{1}=0$, $J_{2}=-1$.

This work is supported by Russian Foundation for Basic Research, grant
13-02-00909a.

\end{document}